# Effect of Ti$_2$Pd(Ni) on the Transformation Behavior in Sputtered Ti-rich TiNiPd Shape Memory Alloys

L. Bumke[1], N. Wolff [2], L. Kienle[2] and E. Quandt[1]

[1] Inorganic Functional Materials, Institute for Materials Science, Faculty of Engineering, Kiel University, Kiel, Germany

[2] Synthesis and Real Structure, Institute for Materials Science, Faculty of Engineering, Kiel University, Kiel, Germany

## Abstract

TiNiPd based shape memory alloys (SMAs) share similar microstructural features as TiNiCu-based SMAs known for their exceptional resistance to functional fatigue due to their high crystallographic compatibility, nanometer sized grains and coherent precipitates, making them an ideal system to further explore the critical factors influencing cyclic stability. In this study, we investigate the effect of heat treatments (500 °C, 600 °C, 700 °C and 800 °C) on the cyclic stability and microstructure of free-standing, magnetron-sputtered Ti$_{53.6}$Ni$_{35.2}$Pd$_{11.2}$ films. All heat treatments promote the formation of Ti$_2$Pd(Ni) precipitates and result in a similar grain size (~1-4 µm). Lower heat treatment temperatures improve the cyclic stability of the stress induced transformation while reducing transformation temperatures and latent heat. Temperature dependent X-ray diffraction reveals a complex microstructure for the martensite phase with Ti$_2$Pd(Ni), Ti$_2$Ni(Pd), TiNiPd(B2), B19/B19' and R-phase. The thermal phase transition changes from a distinct 1$^{st}$ order to a 2$^{nd}$ order like transition, accompanied by increasing amount of remanent austenite and R-phase, with nearly no change for the sample heat treated at 500 °C. *In situ* stress dependent X-ray diffraction demonstrates a significant difference between the temperature and stress induced phase transformation for this heat treatment. The observed semi crystalline microstructure, featuring nano domains of Ti$_2$Pd(Ni) precipitates in the sample heat-treated at 500 °C, leads to a mixture of long range martensitic and strain glass transition. This study highlights the impact of heat treatment and microstructure on the phase transformation behavior and functional fatigue in Ti-rich TiNiPd alloys.

## Introduction

In the recent years significant progress has been made in improving the functional fatigue of TiNi based shape memory alloys through precipitation [1, 2], grain size [3–5] and phase engineering [6–8]. Functional fatigue describes the ability of a material to maintain the solid to solid 1$^{st}$ order phase transformation without degradation that would otherwise result in a diminishing effect size [9]. Among SMAs, TiNiCu alloys have shown exceptional functional stability by combining high crystallographic compatibility, nanometer sized grains and coherent tetragonal precipitates [6, 10]. Separating the individual contributions of these factors to functional fatigue remains a challenge, particularly in polycrystalline materials. Crystallographic compatibility is crucial to control thermal hysteresis and potentially reduces the formation of dislocation at the interface of martensite and austenite. The crystallographic compatibility can be described by the cofactor conditions (CC) consisting of three conditions [8, 11, 12]. The first conditions described the misfit of the austenite and martensite lattice at the interface and if $\lambda_2 = 1$ (ordered middle eigenvalue of the transformation stretch matrix ***U***) no elastic energy is stored at the interface formed by austenite and a single variant of martensite. The fulfilment of the other two conditions (CCII and CCIII) allows for the formation of an infinite number of zero stress interface of austenite with twinned martensite and thus enhancing fatigue resistance [13]. In TiNiCu alloys, the crystallographic compatibility gradually improves with increasing Cu content and approaches $\lambda_2 = 1$ (stays below 1), [14] resulting in low thermal hysteresis and improved fatigue resistance. In addition, grain size decreases to the sub µm range with increasing Ti and Cu content [2, 15], while Guinier-Preston (GP) zone like and coherent Ti$_2$Cu(Ni) precipitates significantly improve functional fatigue resistance [2, 10, 16]. This have led to the identification of a wide compositional range with negligible functional fatigue of superelasticity even after millions of cycles. It would be desirable to transfer this microstructural engineering to other material system, to create materials with a similar resistance to functional fatigue. TiNiPd alloys, widely studied as high temperature shape memory alloys, share some similarities with TiNiCu including:

- B2-B19 phase transition for Pd larger 7-10 at.% [17, 18]
- High crystallographic compatibility in the range of 9-11 at.% Pd, although with worse CCII values and strong compositional dependence of the compatibility especially for lower Pd content [14].
- Coherent tetragonal (I4/mmm) $Ti_2Pd(Ni)$ precipitates with the same orientation relationship as $Ti_2Cu(Ni)$/GP zones with the matrix [19, 20].

Despite these similarities, TiNiPd alloys exhibit notable differences. The transformation temperatures are highly sensitive to the Ti and Pd content in the matrix. A decrease in Ti content below 50 at.% leads to a significant reduction in transformation temperature [21–23] and ultimately can lead to the formation of a strain glass state at $Ti_{49}Ni_{51-x}Pd_x$ (x=7.5-15 at.%) caused by the point defects introduced by a surplus of Ni and Pd atoms [24]. In contrast TiNiCu alloys show only a moderate decrease for similar deviations from 50 at.% (above or below) [2] and strain glass behavior is induced at much higher Ni defect doping $Ti_{50-x}Ni_{35+x}Cu_{15}$ (x=7-9 at.%) [25]. Increasing the Pd content above 10 at.% results in a pronounced rise in transformation temperature, while TiNiCu alloys exhibit only a minor increase with higher Cu content [17, 26]. The lattice misfit between $Ti_2Pd(Ni)$ precipitates and the matrix is typically around 1-3% [19, 20, 27], ensuring coherent growth even at larger precipitate sizes. In comparison, $Ti_2Cu(Ni)$ precipitates exhibit a misfit of approximately 3%, leading to reduced coherency at larger sizes [16]. While the positive effect of $Ti_2Pd(Ni)$ precipitates to suppress dislocation formation is well documented [19], most studies have focused on systems with high Pd content. Research on sputtered TiNiPd films has mainly explored high-temperature SMAs [19, 28]and their microstructure, thermal hysteresis and constant force thermal cycling [18], whereas functional fatigue of the superelasticity remains largely unexplored [29], particularly for Pd concentrations below 15 at.%, where the crystallographic compatibility is superior compared to the high temperature SMAs.

In this study we investigate the influence of $Ti_2Pd(Ni)$ precipitates on the functional fatigue of the stress induced phase transition in sputtered $Ti_{53.6}Ni_{35.2}Pd_{11.2}$ SMAs. These alloys are in a compositional range where a high crystallographic compatibility [14], as well es $Ti_2Pd(Ni)$ precipitates are present [30] both of which are known to enhance the resistance to functional fatigue. The effect of different heat treatments on the microstructure and the influence on both stress induced and thermally induced phase transitions will be investigated. Temperature and stress dependent X-ray diffraction together with transmission electron microscopy (TEM), will be used to clarify the role of microstructure on the martensitic transition and in improving the cyclic stability of TiNiPd alloys.

**Experimental**

Amorphous Ti-rich TiNiPd films with a thickness of approximately 15 µm were deposited on a 4-inch structured Si wafer using a Von Ardenne CS730S (Von Ardenne, Germany, base pressure < $3x10^{-7}$) cluster magnetron sputtering system with a 4-inch TiNiPd target.The films were deposited by dc magnetron sputtering in a planar configuration. The structured films were released from the substrate by sacrificial layer etching [31]. A compositional mapping over the 4-inch Si wafer of the amorphous film, was measured using a Helios Nanolab 600 scanning electron microscope (SEM) (FEI, Germany) equipped with an energy dispersive X-ray spectroscopy (EDS) silicon drift detector (Oxford Instruments, UK). A $Ti_{50}Ni_{40}Pd_{10}$ standard measured by inductively coupled plasma optical emission spectrometry (ICP-OES) was used to increase the accuracy of the measurement. The average film composition was then determined by averaging over the complete 4-inch wafer area. The structured films consist of ring dogbones for mechanical testing, circles for DSC measurements and rectangular samples for X-ray diffraction (XRD) measurements (See Fig. S1, Supporting Information (SI)). For crystallization, the TiNiPd films were heat treated using a rapid thermal annealing (RTA) system, RTA-6 SY09 (Createc Fischer, Germany) at 500 °C, 600 °C, 700 °C and 800 °C for 15 minutes. Transition temperatures were measured by differential scanning calorimetry using a 204 F1 Phoenix (Netzsch, Germany) in the range of +120 °C to -120 °C.

The mechanical properties were determined using a Z.05 universal tensile tester (Zwick Roell, Germany). The machine is equipped with a thermal chamber. The strain rate was set at 2%/min to ensure isothermal conditions with minimal self-heating and cooling. The ring dogbone shaped samples had a parallel length of 8 mm (clamping length ~ 5.45 mm) and a parallel width of 0.5 mm. The samples were tested at temperatures close to $A_f$ and to a maximum stress of 500 MPa for the sample annealed at 800 °C, 700 MPa for the sample heat treated at 600 °C and 700 °C and to 800 MPa for the sample annealed at 500 °C. The sample annealed at 500 °C is additionally electropolished, due to its increased brittleness compared to the other samples. The strain was corrected in such a way for each cycle that the maximum strain aligns with the maximum strain of the 1st cycle to reduce the effect of temperature variation of the traverse and load cell which are placed outside of the temperature chamber. The fatigue testing was stopped at cycle 200 or when the sample fractured in advance.

The cantilever deflection method was used to analyze the transformation characteristics of TiNiPd films heat treated at 500 °C, 600 °C and 700 °C. Films with a thickness $t_f$ ~ 3-4 µm were deposited onto Mo substrates with a thickness $t_s$ ~ 100 µm. The temperature dependent film stress of the clamped cantilever with a free length $l$ is determined from its deflection $d$, measured via a laser beam reflected at its free end. Heating and cooling between -50 °C to 80 °C is realized by a Peltier Element with a heating and cooling rate of approximately 10 °C/min. The film induced stress is given by the Stoney equation [32] with $\sigma = \frac{1}{3l^2}\frac{E_s}{(1-v_s)}\frac{t_f}{t_s^2}d$ where $\frac{E_s}{(1-v_s)}$ represents the biaxial modulus of the substrate, with $E_s$ as the Young's modulus and $v_s$ as the Posson's ratio of the substrate.

XRD patterns were collected using a SmartLab 9 kW diffractometer (Rigaku, Japan) equipped with a CuKα radiation source (λ = 1.5406 Å) and a 2D Hypix3000 detector operating in 1D mode. A DCS350 heating and cooling stage (Anton Paar, Germany) was used for lattice parameter determination at different temperatures. A 3D printed dome with a polyimide foil and an integrated beam knife, was used as an enclosure under vacuum to avoid condensation and stray radiation. Lattice parameters were determined using the PDXL software (Rigaku, Japan), based on peak positions only. A split pseudo-Voigt function was used for peak fitting as integrated into the software used for indexing. To gain further insights into the phase transformation kinetics, *in situ* transmission mode tensile XRD experiments were performed on the sample heat treated at 500 °C using the same diffractometer in 2D mode in combination with a home-made tensile test holder up to stress to 860 MPa. A parallel X-ray beam in combination with a 300 µm collimator was used to place a beam spot at the center of the sample. The sample holder was rotated 90 degrees (chi = 90°) to reveal the lattice parameter perpendicular (chi = 0°) and parallel to the stress direction (chi = 90°). A range of 17.5° to 47° was examined. The 2D pattern was converted into a line profile by cake integration of 30° slices using 2DP software (Rigaku).

The grain structure was revealed using a HF based solution. Images were taken by SEM using a Gemini Ultra55 Plus (Zeiss, Germany) equipped with an SE detector.

In extension of averaging XRD analysis, selected samples of the TiNiPd alloy were subjected to transmission electron microscopy (TEM) examination. Here, selected-area electron diffraction (SAED) experiments using a Tecnai F30 G$^2$ STWIN microscope were conducted for phase analyses of the shape memory alloy microstructure after annealing at 800 °C. Simulation of electron diffraction patterns was performed using Pierre Stadelmann's JEMS electron microscopy simulation software [33].

Local chemical information about the specimen was retrieved from STEM energy-dispersive X-ray spectroscopy (EDS) experiments conducted on a probe-corrected JEOL NEOARM microscope (200 kV, coldFEG) equipped with a dual silicon drift detector (SDD) system (each 100 mm$^2$ active area). Preparation of cross-section specimens was performed using the standard focused-ion-beam (FIB) technique on a Helios Dual Beam FIB-SEM.

## Results

**Composition and temperature induced transformation**

The average film composition was determined to be $Ti_{53.6}Ni_{35.2}Pd_{11.2}$ across a 4-inch wafer (see Fig. S1, SI). The crystallization temperature was determined to be approximately 496 °C (see Fig. S2, SI) which is in line with results from Kim et al. [34]. The amorphous film was subsequently heat treated at 500 °C, 600 °C, 700 °C and 800 °C for 15 minutes each to induce crystallization. Residual stress was observed in the film heat treated at 500 °C. Transformation temperatures were measured by DSC over a temperature range from -120 °C to 120 °C (Fig.1a-d). The films heat treated at 800 °C (Fig.1a) exhibit a single distinct peak during heating and cooling. At 700 °C (Fig.1b), the peak broadens significantly and displays a shoulder. Annealing at 600 °C (Fig.1c) further reduces the enthalpy, with a small shoulder visible upon heating, however, the DSC signal during cooling is too weak to extract reliable data. In contrast, films annealed at 500 °C do not exhibit any distinct peak but only indications of a broad bump during both heating and cooling (Fig.1d).

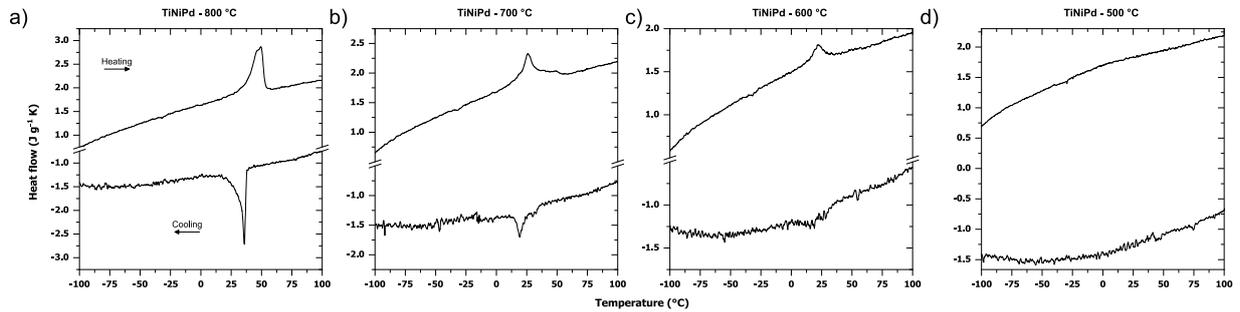

*Figure 1 presents the DSC measurement results for samples annealed at 800 °C (a), 700 °C (b), 600 °C (c) and 500 °C (d). Distinct transformation peaks are observed for the sample heat treated at 600 °C. Note that the increased noise during the cooling cycle is due to liquid nitrogen purging.*

The transformation temperatures and enthalpies, if determinable, are summarized in Table 1. For simplicity, thermal hysteresis is defined as the difference between the heating and cooling peak temperatures, neglecting the additional shoulders.

*Table 1 summarizes the phase transformation temperatures, transformation enthalpies, and thermal hysteresis of the different samples. For simplicity, thermal hysteresis was determined as the difference between the heating and cooling peak temperatures.*

| Heat treatment (°C) | $A_s$ (°C) | $A_p$ (°C) | $A_f$ (°C) | $\Delta H$ (J/g) | $M_s$ (°C) | $M_p$ (°C) | $M_f$ (°C) | $\Delta H$ (J/g) | $\Delta T$ (°C) |
|---|---|---|---|---|---|---|---|---|---|
| 500 | - | - | - | - | - | - | - | - | - |
| 600 | (12.9) | 21.9 | (32.9) | 3.1 | - | - | - | - | - |
| 700 | 21.8 | 26.2 | 55.0 | 8.0 | 25.3 | 19.1 | 13.4 | | 7.3 |
| 800 | 39.7 | 49.5 | 53.1 | 9.8 | 37.7 | 35.9 | 33.3 | -10.0 | 13.6 |

Thermal hysteresis is significantly reduced at lower heat treatment temperatures, also due to the change from a 1-step to a 2-step transition as indicated by the shoulder during cooling and heating. The cantilever deflection method (CDM) was additionally performed for the films annealed at 500 °C, 600 °C and 700 °C as this method is more sensitive to minor transformations or continuous phase changes (see Fig. S3, SI). For these experiments, a ~3.5 µm film was deposited onto a 100 µm Mo substrate. Here no large change in thermal hysteresis was observed for the sample heat treated at 600 °C and 700 °C. A small continuous change of stress together with a vanishing thermal hysteresis was observed in the films heat treated at 500 °C, suggesting $A_f$ of approximately 30 °C. However, the phase transition was not completed at -50 °C (lower limit of the test setup).

**Superelasticity**

The stress-strain curves (Fig. 2) were recorded at approximately 5 °C above $A_f$ for the samples heat-treated at 800 °C (Fig. 2a) and 700 °C (Fig. 2b).

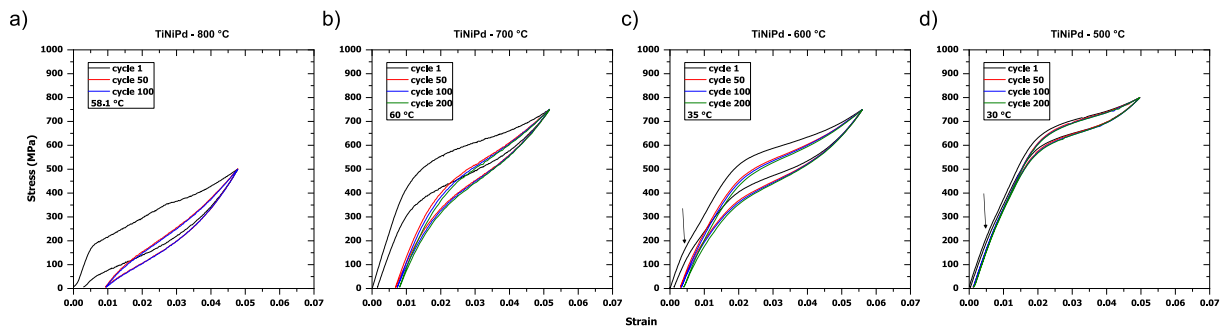

*Figure 2 illustrates the stress–strain curves for TiNiPd shape memory alloy films annealed at 800 °C (a), 700 °C (b), 600 °C (c), and 500 °C (d) over up to 200 superelastic cycles. Note that the slope of the elastic region changes for the sample in (c) and (d) as marked with an arrow.*

In contrast, the 600 °C (Fig. 2c) and 500 °C (Fig. 2d) samples were tested at or near $A_f$ due to their higher transformation stress. For the 500 °C sample, the $A_f$ temperature determined from the CDM experiment was used as the testing temperature. Additionally, this sample required electropolishing due to its increased brittleness compared to the other samples.

The critical transformation stress increases as the heat treatment temperature decreases and functional fatigue is significantly reduced at lower heat treatment temperatures, even though overall stresses are much higher. However, high fatigue testing was not possible due to the severe brittleness observed in these samples, consistent with previous reports of a similar composition [30]. Notably, all superelastic curves exhibit an inclined transformation plateau, with the 800 °C sample showing the steepest slope. The 500 °C sample showed a reduced stress hysteresis to 60 MPa compared to ~ 120 MPa for the 600 °C and 700 °C sample and 150 MPa for the 800 °C sample. Additionally, the samples heat treated at 500 °C and 600 °C both exhibit a change in slope during elastic loading, characteristic for the R-phase transition. Variation in ambient temperature reveals that the onset of this slope change is temperature dependent. Furthermore, the 500 °C sample maintains superelasticity down to 0 °C (Fig. S4, SI).

### *In situ* temperature dependent XRD

Temperature dependent XRD provides additional insights into the thermally induced transformation and the presence of the corresponding phases (Fig. 3).

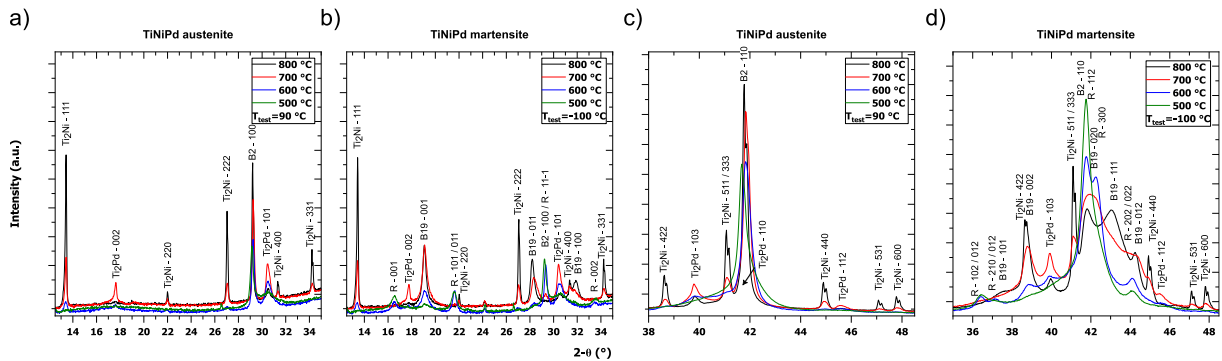

*Figure 3 presents the XRD data in the low-angle region for different samples, showing the austenite phase at 90 °C (a) and the martensite phase at -100 °C (b). The data corresponding to the regions around the 110 B2 reflections are displayed in (c) for the austenite state and in (d) for the martensite state.*

The diffractograms in the austenite state (Fig. 3a, c) tested at 90 °C revealed the presence of a complex microstructure. Reflections corresponding to $Ti_2Pd(Ni)$ precipitates were detected in all heat-treated samples, with the highest intensity observed in the sample annealed at 700 °C. The intensity of $Ti_2Pd(Ni)$ reflections increased with annealing temperature up to 700 °C but decreased at 800 °C. No distinct $Ti_2Ni(Pd)$ reflections were detected in the 500 °C sample. However, the presence of $Ti_2Ni(Pd)$ precipitates increased with higher annealing temperatures. Significant peak broadening was observed for all phases at lower heat treatment temperatures, with strong asymmetric broadening particularly for the 110 reflection in the film heat-treated at 500 °C. At 90 °C, the largest B2 lattice parameter (~3.0618 Å) was measured for the 500 °C sample. The values remained nearly constant for the samples heat treated at 600 °C (3.0528 Å) and 700 °C (3.0518 Å), then increased again to 3.0562 Å for the 800 °C sample. For the $Ti_2Pd(Ni)$ precipitates in the sample annealed at 700 °C (highest intensity), the lattice parameters were determined as $a_{Ti2Pd(Ni)}$ ~ 3.061 Å and $c_{Ti2Pd(Ni)}$ ~ 10.05 Å. The $Ti_2Ni(Pd)$ phase exhibited a lattice parameter of approximately $a_{Ti2Ni(Pd)}$ ~ 11.41 Å. Notably, a $Ti_2Ni(Pd)$ phase with slightly smaller lattice parameter ($a_{Ti2Ni(Pd)}$~11.39 Å), similar to [16], was also detected. A complex microstructure evolves in the martensite phase (Fig. 3b, d). In all cases, remanent austenite is observed down to -100 °C, with its amount increasing as the heat treatment temperature decreases. The samples heat treated at 500 °C and 600 °C additionally show the R-phase (without distinct B2 110 splitting) and B19/B19', while the 700 °C and 800 °C samples only show reflections corresponding to B19/B19' besides remanent B2. It is not clear whether the martensite phase is monoclinic or orthorhombic, as peak broadening and the presence of precipitates hinder precise structural determination. Literature suggests that a "low-angle" martensite phase (β < 91°) evolves at lower temperatures following the formation of the orthorhombic phase [35, 36]. To further investigate the phase transformation kinetics, cooling and heating experiments were conducted (see Fig. S5–S8, SI). The following observations were made for the different samples. In all cases the onset of martensitic transition agrees well with the DSC or CDM result. For the 500 °C sample a R-phase transition occurs first, followed by a small simultaneous or subsequent B19/B19' transformation. At -100 °C, a substantial amount of remanent austenite remains. The 600 °C sample exhibits a behavior similar to that of the 500 °C sample. The R-phase transition occurs alongside the B19/B19' transformation, but without the prominent B2 110 reflection splitting typically observed in binary NiTi. In this case, a larger fraction of the material transforms into the B19/B19' phase. For the 700 °C sample an R-phase transition occurs together with the B19/B19' transformation. However, the R-phase transition is completed at approximately -25 °C, after which the R-phase nearly completely diminishes in favor of the orthorhombic/monoclinic phase. Additionally, the amount of remanent austenite is further reduced. In contrast to the other samples, no R-phase transition is observed in the 800 °C sample. Instead, the transformation appears to be more distinctly 1$^{st}$ order and remanent austenite, indicated by the B2 100 reflection is nearly undetectable. Changes in the martensitic structure persist down to -100 °C. Still, in all cases, a more continuous phase transition occurs, leading to a gradual change in lattice parameters for the martensite phase. Accurate lattice parameter refinement was therefore not possible. However, for completeness, the lattice parameters of the R-phase were determined for the 600 °C heat treated sample based on the 001 and 101 R-phase reflections, resulting in approximately $a_{R-phase}$ ~ 7.35 Å and $c_{R-Phase}$ ~ 5.365 Å. The lattice parameters of the martensite phase, assuming an orthorhombic structure, were determined for the 800 °C heat treated sample using the 001, 011 and 100 reflections, yielding $a_{B19}$ ~ 2.805 Å, $b_{B19}$ ~ 4.311 Å and $c_{B19}$ ~ 4.655 Å. We also calculated the ordered middle eigenvalue ($\lambda_2 = b_{B19}(\sqrt{2} * a_{B2})$) of the transformation stretch matrix (U). Using the lattice parameters at 90 °C and -100 °C for the 800 °C heat-treated sample, we obtain $\lambda_2$ = 0.9974. Additionally, using the lattice parameters at 25 °C and 45 °C (closer to phase coexistence) for austenite ($a_{B2}$ ~ 3.0525 Å) and martensite ($b_{B19}$ = 4.3164 Å), $\lambda_2$ was determined to be ~ 0.9998. These values should be interpreted with caution, as they were derived from single reflections, not a whole powder pattern approach and may be influenced by the continuous change in lattice parameters, caused not only by thermal expansion but by gradual lattice changes by the phase transition. Nevertheless, the calculated $\lambda_2$ values indicate an average high degree of crystallographic compatibility, suggesting potentially good fatigue resistance.

*In situ* **stress dependent XRD**

The temperature dependent XRD and DSC measurements indicate that a complete phase transition does not occur within the samples up to 700 °C and follows a continuous rather than discrete transformation behavior. In contrast, tensile tests (Fig. 2) show a distinct phase transition with similar transformation strain of several percent for all heat treatments and a clear sign of R-phase transition for the samples heat treated at 500 °C and 600 °C. To better understand the microstructural changes during the stress induced phase transformation compared to the thermally induced transformation, *in situ* XRD studies were performed in transmission geometry during tensile loading up to ~ 860 MPa on a sample heat treated at 500 °C (Fig. 4). The resulting diffractograms for reflections parallel (∥) and perpendicular (⊥) to the stress direction at different stress values, recorded in the 2θ range of 17.5°-47.0°, are shown in Fig. 4(a-d).

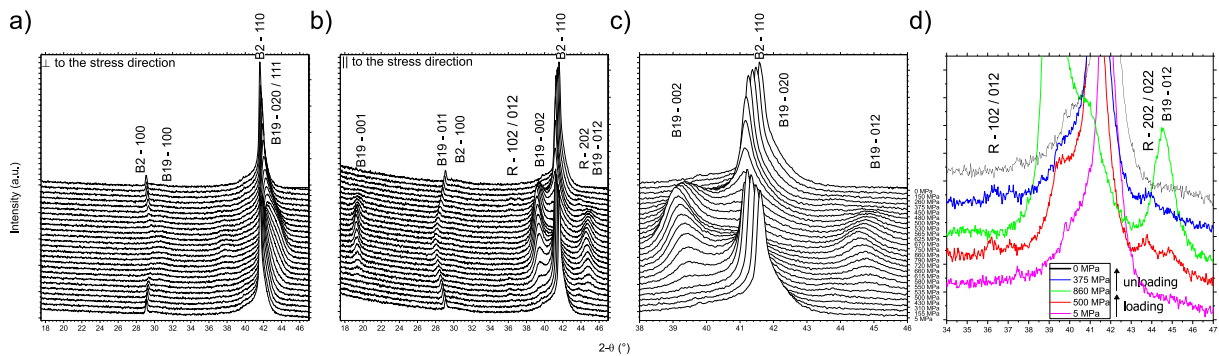

*Figure 4 displays the resulting X-ray diffraction data of the sample heat treated at 500 °C perpendicular to the stress direction (a) and parallel to the stress direction (b). In (c) the waterfall plot of (b) for the region around the B2-110 reflections is enlarged. In (d) selected stress states are represented to visualize the R-phase transition occurring in this sample.*

During elastic deformation, the perpendicular lattice planes experience compressive stress (Fig. 4a), while those parallel to the stress experience tensile stress (Figs. 4b-d). For both orientations, the initial elastic strain or contraction is followed by a phase transformation to the textured R-phase at about 310 MPa parallel and 155 MPa perpendicular to the stress direction. At these stress levels, additional weak reflections corresponding to the R-phase emerge (Fig. 4d). In the perpendicular case, this is only visible as a decrease in the intensity of the B2 110 reflection. With increasing stress, reflections of B19 or B19' martensite appear. Perpendicular to the stress direction are the 020 and 111 B19 reflections and possibly weak 001 and 101 B19 reflections. However, the positions of the reflections do not exactly match the expected values for either the R phase or the B19 phase. In contrast, parallel to the direction of stress, the phase transformation is characterized by a decrease in the intensity of the 110 reflections along with the appearance of highly textured martensite reflections, particularly the 001, 011, 002, and 012 reflections. No indication of the 111 reflection is visible. As the stress continues to increase, the R-phase reflections gradually disappear, leaving only those corresponding to B19/B19' martensite. Upon unloading, the phase transformation is completely reversed to the austenite state: the R-phase reflections reappear during unloading and then disappear completely when the sample is completely unloaded (Fig. 4d) and no residual martensite or peak broadening, commonly observed in binary NiTi, is detected. Compared to the temperature induced transformation, a distinct B2→R phase→B19/B19' transition is observed under stress. Though the transititon e.g. from 100 B2 to 011 B19 is more continuous. The transformation is nearly complete at a stress level of 860 MPa, as indicated by only small changes in the intensities of the B19/B19' reflections. It is generally expected that not well aligned grains will transform at a much later stage [37]

**Microstructure**

The grain sizes for all heat treatments were revealed by wet chemical etching and determined to be in the range of 1-4 µm (see Fig. S9, SI), with a trend toward smaller grain sizes (1-2 µm) for the high temperature heat treated samples. TEM was performed on samples annealed at 500 °C and 800 °C. The microstructure of the sample annealed at 800 °C is presented in Fig. 5 and contains grain boundary precipitates of $Ti_2Pd(Ni)$ and $Ti_2Ni(Pd)$. Within the grains, rod-like $Ti_2Pd(Ni)$ precipitates and round $Ti_2Ni(Pd)$ precipitates are observed. According to Matsuda et al. [20] the appearance of rod like precipitates can be rationalized by the difference in coherency strains depending on the direction. Taking the lattice parameters as determined by XRD the coherency strain in our case is $\left|\frac{d(003)_{Ti_2Pd(Ni)} - d(100)_{TiNiPd}}{d(100)_{TiNiPd}}\right| * 100 = 9.8\%$, and $\left|\frac{d(100)_{Ti_2Pd(Ni)} - d(100)_{TiNiPd}}{d(100)_{TiNiPd}}\right| * 100 = 0.3\%$, persevering coherent growth to larger sizes of $Ti_2Pd(Ni)$ precipitates. The sample is initially in the martensite state but transforms into austenite upon electron beam heating. Here $Ti_2Pd(Ni)$ precipitates impede austenite growth, causing some regions to transform earlier than others (Fig. S10, SI). The rod-like $Ti_2Pd(Ni)$ precipitates are found to be coherent with the matrix which is demonstrated by the SAED pattern recorded along the [110] direction of the $Ti_2Pd(Ni)$ and TiNiPd-B2 phases (Fig. 5b). Simulated superposition patterns (Fig. 5c) reveal that such coherency can only be achieved if the $Ti_2Pd(Ni)$ c-axis is locally expanded by approximately 7%, reaching about 10.8 Å, in contrast to the relaxed lattice parameter determined by X-ray diffraction with c-axis value of ~ 10.05 Å. Similar discrepancies have also been reported by P. Schlossmacher [30]. Within the investigated area, $Ti_2Pd(Ni)$ precipitates with different lattice parameters were found (Fig. S11), which agrees with the broad reflections visible in the XRD. The lattice parameters determined for the TiNiPd-B2 phase (3.051 Å) and $Ti_2Ni(Pd)$ (11.38 Å) agree with the values extracted from the X-ray diffraction data. Figure 5b and 5d depict SAED patterns along the [110] and [100] directions, indicating that the (002) planes of $Ti_2Pd(Ni)$ are aligned parallel to the (001) planes of the matrix with coherent intergrowth on (1-10) and (020) planes, respectively. Based on the SAED analysis and in agreement with the literature, the following orientation relationship can be established (001) B2||(002) $Ti_2Pd(Ni)$. $Ti_2Ni(Pd)$ precipitates inside the grain are coherent or semi-coherent with the matrix, as evident by the appearance of speckled intensities around main structure reflections in Fig. 5d and the formation of Moiré fringes on intergranular $Ti_2Ni(Pd)$ caused by the difference in lattice spacing with TiNiPd B2 (Fig. S11, SI). The misfit of the $Ti_2Ni(Pd)$ precipitates with the matrix are approximately $(100)B2||(100)Ti_2Ni(Pd) = \frac{4c(B2)}{a(Ti_2Ni(Pd))} \approx$ 6.95%. Along the grain boundaries the $Ti_2Ni(Pd)$ precipitates are observed in multiple orientations (e.g., [111], [110] and [100]) (Fig. 5e), whereas within a [110] oriented grain only precipitates in [101] orientation are evidenced. TiNiPd-based shape memory alloys are highly sensitive to the Pd content, which significantly influences the transformation temperatures and thermal hysteresis. Similarly, the Ti content plays a critical role, comparable to the behavior observed in binary NiTi alloys. Typically, when the Ti content is below 50 at.%, a steep decrease in transformation temperatures is observed [19, 21, 23]. However, in all our cases, the Ti composition in the matrix was found to be close to or below 48 at.% by TEM-EDS, which is significantly lower than the measured value of the as-sputtered and not-annealed film by SEM-EDS.

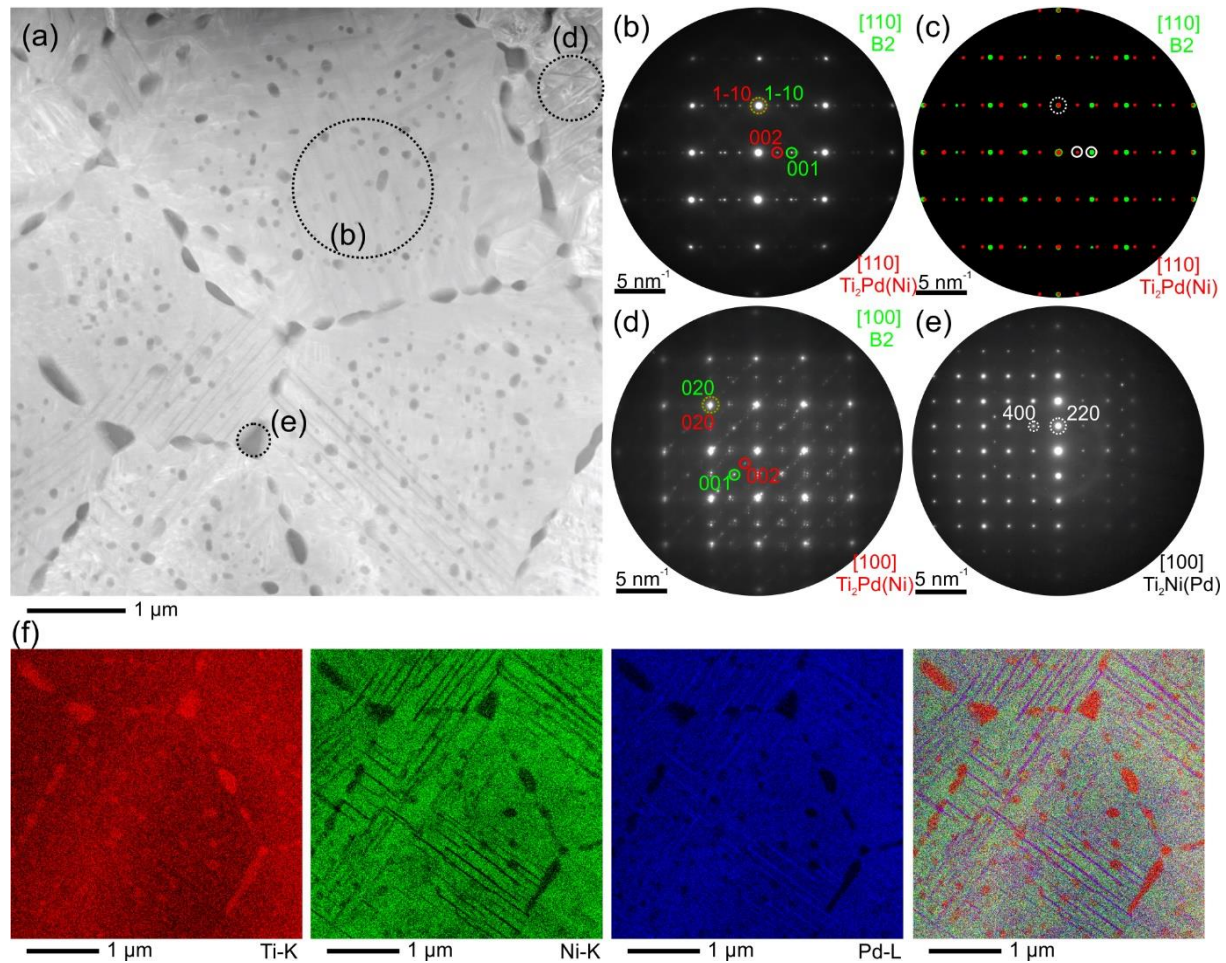

*Figure 5 Structure analysis by TEM. (a) STEM-HAADF micrograph of the TiNiPd (800 °C) microstructure exhibiting grain boundary precipitates of $Ti_2Ni(Pd)$ and intergranular round $Ti_2Ni(Pd)$ and rod-like $Ti_2Pd(Ni)$ precipitates. (b) SAED pattern from the center region of a large B2-phase TiNiPd alloy grain composed of [110] B2 phase and [110] $Ti_2Pd(Ni)$ zone axis patterns. (c) Superposition of the calculated patterns of the components shown in (b) demonstrates the coherency between B2 phase and rod-like $Ti_2Pd(Ni)$ precipitates. (d) SAED pattern showing the structural coherency of B2 and $Ti_2Pd(Ni)$ in [100] orientation. Speckled intensities around main structure reflections arise from coherent $Ti_2Ni(Pd)$ grains. (e) [100] zone axis SAED pattern of one $Ti_2Ni(Pd)$ grain boundary precipitate .(f) Single elemental maps of Ti-K (red), Ni-K (green), Pd-L (blue) and superposition image recorded by EDS.*

Similar low Ti-values were reported by [38], which contrasts the results of Delville and Schryvers [27]. This discrepancy needs further clarification. Therefore, the reported compositions are used for qualitative discussion and relative comparisons rather than as absolute values. The overlay of the EDS map in Fig. 5f, illustrates a noticeable compositional variation within the grain. The overall composition of the matrix was determined to be $Ti_{48.0}Ni_{39.2}Pd_{12.8}$ at.%, which is significantly different from the composition determined by SEM-EDS for the amorphous film, as discussed above. Detailed STEM-EDS maps can be found in the SI (Fig. S12 and Table S1, SI). Inside the grains, the composition of $Ti_2Ni(Pd)$ precipitates is more variable and is enriched in Ti (55.7±1.9 at.%) while they contain less Ni (33.9±0.8 at.%) and Pd (10.5±1.2 at.%), compared to the $Ti_2Ni(Pd)$ (Ti: 62.6±0.2; Ni: 30.7±0.1; Pd: 6.7±0.1 at.%) puckered along the grain boundaries. Next to the GB $Ti_2Ni(Pd)$ precipitates, occasionally, round GB $Ti_2Pd(Ni)$ were observed (Fig. S13), which show about 3 at.% higher Ti content compared to intergranular precipitates, while exhibiting a similar Pd content. Overall, the stoichiometry of the TiNiPd matrix between the intergranular $Ti_2Pd(Ni)$ precipitates is more deficient (~3 at.%) in Ti and enriched in Ni compared to the matrix around the intergranular $Ti_2Ni(Pd)$ precipitates. Depending on the location, variations in the matrix composition of Ti and Ni range from 1 to 2.5 at.%, with the Pd content varies around 1 at.%. The microstructure of the sample heat treated at 500 °C is shown in Fig. 6a and includes regions that are mostly amorphous or semicrystalline with small rod-like precipitates (Fig. 6b), crystalline regions without visible precipitates and regions that show complete crystallization with cross-shaped precipitates (Fig. 6c). These intergranular cross-shaped precipitates have an extremely small size of ~1 nm and could be identified as $Ti_2Pd(Ni)$ precipitates which seem to form first from the amorphous matrix.

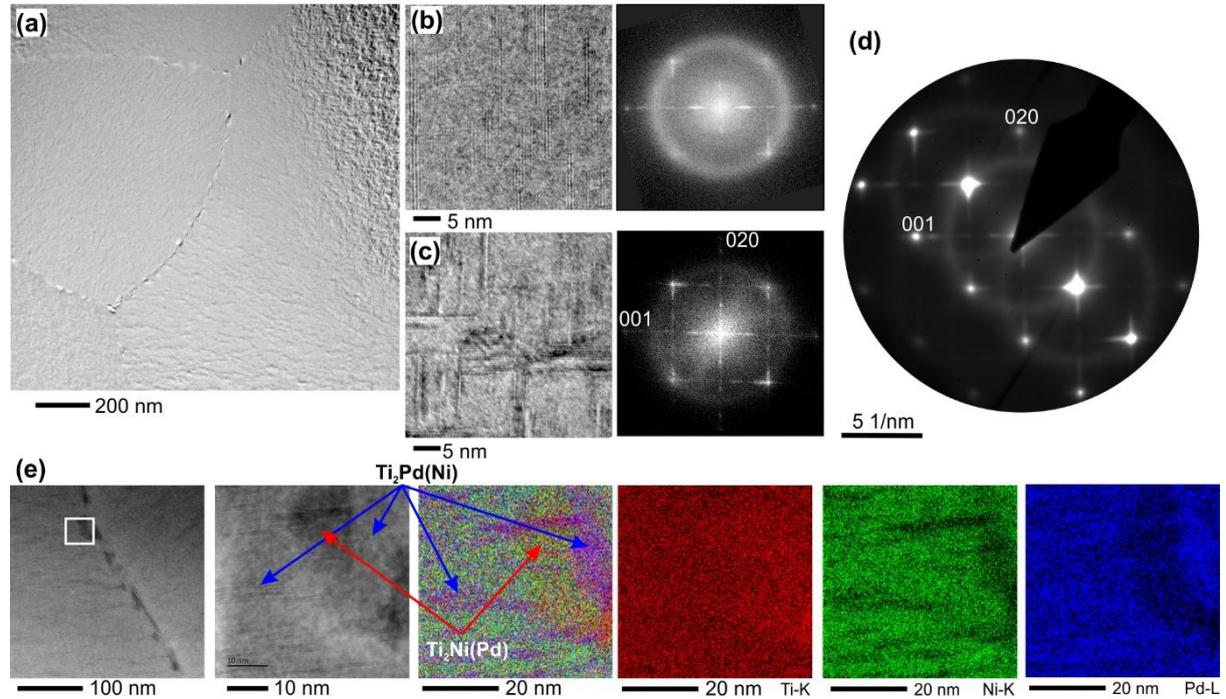

*Figure 6 Structure analysis by TEM. (a) STEM-HAADF micrograph of TiNiPd (500 °C) microstructure with small grain boundary precipitates. The matrix consists of an amorphous/semicrystalline matrix with local areas being less and more crystalline (b) Shows the semicrystalline matrix together with thin rod precipitates. The FFT shows a pronounced ring of diffuse intensity corresponding to amorphous material and reflections belonging to the [100]B2 phase . (c) shows the crystallized matrix with cross-shaped rod-like precipitates. The inset shows the corresponding FFT. In (d) a SAED pattern of a region wth cross-shaped precipitates is presented. The coherent growth of $Ti_2Pd(Ni)$ is evident from diffuse streaks along {001}\*. (e) Examination of precipitates at a grain boundary. Single elemental maps of Ti-K (red), Ni-K (green), Pd-L (blue) and an EDS superposition image of a region at the grain boundary are shown. Arrows indicate the presence of $Ti_2Pd(Ni)$ and probably $Ti_2Ni(Pd)$ precipitates.*

The SAED pattern of the TiNiPd phase along the [100] direction in Figure 6d shows diffuse streaks along the [001]\* and [020]\* directions originating from the $Ti_2Pd(Ni)$ precipitates and confirming the coherence to the matrix [19, 27]. In most cases, the rod-like precipitates in the sample annealed at 800 °C showed one preferred growth direction within a common area of the grains, which is in contrast to the cross-shaped pattern observed for the 500 °C sample. Hence, it is assumed that precipitate annihilation and growth of one orientation takes place during further annealing. In addition, small precipitates are already found along the grain boundaries consisting of presumably $Ti_2Ni(Pd)$ and $Ti_2Pd(Ni)$ precipitates (Fig. 6e) as indicated by STEM-EDS and confirm that the precipitates form in the early stage of crystallization.

## Discussion

Our results reveal a strong influence of the heat treatment temperature on the shape memory behavior. In addition, we find a pronounced difference between the thermal and stress induced phase transition in the sample heat treated at 500 °C. We believe, though not yet experimentally confirmed that the difference in stress induced and thermal induced transformation can be likely found as well with smaller magnitude for the sample heat treated at 600 °C and 700 °C. The result and differences can be potentially explained by the formation of $Ti_2Pd(Ni)$ precipitates. The grain size as influencing factor can be excluded, as the difference between the heat treatment is neglectable and the size of the grains is much larger than needed for the suppression of martensitic transition [3]. The coherency strains originating from $Ti_2Ni(Pd)$ precipitates can be neglected as well, as they are believed to be incoherent or semicoherent. The different heat treatment lead in all cases to the formation $Ti_2Pd(Ni)$ precipitates with different size and spacing of the rod like precipitates possibly leading to different contributing internal strains, as well as compositional variations. In case of the 500 °C sample additional defects are present, caused by the partial amorphization of the sputter process. The diminishing intensity of $Ti_2Pd(Ni)$ precipitates at 800 °C indicates, that the precipitates become unstable. The transformation behavior observed especially for the sample heat treated at 500 °C strongly resembles that of a strain glass induced by the formation of $Ti_2Pd$ nanoprecipitates [39]. Strain glasses are a special class of ferroic glasses, characterized by a frozen disorder state of local lattice distortions, experimentally first described by Sarkar et al [40]. They are observed in SMAs where the martensitic phase transition is suppressed by the introduction of a high density of structural defects [41], caused by the introduction of point defects by doping/co doping (compositional variation) [40], nanoprecipitates [39, 42, 43], nanograins [41] and dislocations/cold rolling [44, 45] or a combination of those features. These defects locally distort the lattice and lead to the formation of uncorrelated nano strain domains with only a short-range strain order interrupting the spontaneous, cooperative structural change and the resulting long-range strain ordering of the martensitic transformation. In the view of the free energy landscape, local random orientated strain nanodomains cause the presence of many equivalent local energy minima leading to a frustration within the system[40]. Typically the energetic favorable martensitic state cannot be reached, as local energy barriers are too high and are increasing with decreasing temperature, as the strain glass process is kinetic and not thermodynamically driven [46]. Macroscopically, below the transition temperature the structure is cubic B2 with frozen local disordered nanodomains and the local energy minima can only be overcome by applying an external stress to the system, allowing to reach the energetically stable long range ordered martensite (global minima) configuration [46]. The "superelasticity" in this case is achieved by the reversible stress induced transformation from macroscopically B2 with local strain disordered nanodomains to martensite and contrasts the invariance of the austenite phase when only the ambient temperature is reduced. To achieve the strain glass state in TiNiPd based SMAs different approaches allow for the strain glass formation. Ren et al. [24] showed that the strain glass can be introduced into the system by decreasing the Ti-content to 49 at.% and doping together with the surplus of Ni with Pd leads to the strain glass formation caused by point defects at Pd concentrations between 7.5 and 15 at.%. With higher Pd contents, B19 is stabilized as a competing factor. Introducing $Ti_2Pd$ precipitates together with additional doping of Fe causes the strain glass formation. Strain glass formation by aging however allows for a mixture of strain glass transition and spontaneous martensite transition [39, 47]. The formation of strain glasses can also explain why in the TiNiPd system, the martensitic transition vanishes at Ti-contents < 49 at.% [24]. For instance, A. A. Klopotov et al. [21]found a strain glass like behavior in $Ti_{49}Ni_{43.5}Pd_{7.5}$, but did not link it to strain glasses at that time. The recently proposed reentrant strain glasses [48], potentially explains, why in TiNiPd based alloys the orthorhombic phase is undergoing a small shear upon cooling to low temperatures [36] and why in TiNiPd a substantial amount of non-transforming austenite is observed [49]. Evidence for strain glass behavior is often demonstrated by the frequency dependence of dynamic properties (DMA), broken ergodicity in zero field and field cooling measurements and nanodomain formation without a macroscopic phase change (XRD, HRTEM). The absence of DSC peaks, increasing electrical resistance with decreasing temperature and simultaneous lack of thermal hysteresis are further signatures. In contrast, these materials exhibit still depending on test temperature superelasticity and shape memory effect associated with stress induced martensite formation, as shown by *in situ* XRD. Though we did not explicitly test for all these signatures, there is strong evidence that at least the sample heat treated at 500 °C shows a strain glass like transformation behavior together with some amount of martensitic transformation. We believe that also for the samples heat treated at 600 °C and 700 °C there are areas where martensitic transition is suppressed and undergoes a strain glass like transition. The local variance of transformation behavior is supported by the variation of local microstructure (Fig. 5, 6). A mixture of martensite transition together with strain glass like transition can be expected for those alloys as recently shown also for binary NiTi [47] and TiPdFe [39]. A gradual change with heat treatment temperature is therefore expected.

**Thermally Induced Transformation**

The characteristic of the thermally induced transition is greatly influenced by the reduction of heat treatment temperature. The reduction of latent heat and the change of transformation sequence can be explained by the formation of a complex microstructure as evident by the TEM and XRD results. The sample heat treated at 800 °C shows a clear single step B2 → B19 transition, supported by temperature dependent XRD results, which may also indicate an additional symmetry lowering to B19', a feature not detected by DSC through the continuous change of symmetry. TEM-EDS analysis revealed compositional variations depending on grain location, particularly in the vicinity of $Ti_2Ni(Pd)$ or $Ti_2Pd(Ni)$ precipitates. While the matrix Pd concentration is in a range favoring a direct B2 → B19/B19' transition, the measured Ti content (45.8 - 48.6 at.%) requires further clarification, since such low Ti levels would typically preclude a martensitic transition in TiNiPd alloys, as discussed already above. It is important to stress that the reduced precipitate spacing with decreasing heat treatment temperatures can potentially affect local strains and transformation temperatures. Although the spacing between $Ti_2Pd(Ni)$ precipitates is greater than 100 nm, well above the critical limit for R-phase and B19/B19' formation [3, 50], their presence still slightly hinders the nucleation and growth processes. This explains why the thermal hysteresis in this Ti-rich alloy is greater than in other TiNiPd alloys with compositions close to Ti 50 at.% and $\lambda_2$ close to 1 [8] and why a small but detectable amount of remanent austenite persists down to -100 °C. This remanent austenite is probably caused by compositional fluctuations or coherent stresses in the vicinity of the precipitates [27]. However, the overall influence of coherent stress appears to be small, as evidenced by the minimal remanent austenite upon cooling and with latent heat values, which are consistent with the reported literature [51] considering the substantial amount of secondary phases. Reducing the heat treatment temperature to 700 °C changes the transformation sequence to B2 → B2 + R-phase → B2 + B19/B19', as evidenced by the shoulder in the cooling and heating curve in Fig. 1b and supported by XRD data. Although no TEM data are available, the appearance of the R-phase is likely due to local compositional variations and coherent stresses from $Ti_2Pd(Ni)$, similar to $Ti_3Ni_4$ precipitates [52–54], their reduced spacing or a combination of those factors, which suppresses B19/B19' formation. This results in increased retained austenite and reduced latent heat. A smaller thermal hysteresis (peak to peak) is not caused by an improved crystallographic compatibility, as the matrix composition likely remains highly heterogenous, but stems from the change in transformation sequence via the R-phase. In addition, smaller precipitates may be less of an obstacle to phase transformation, which would favor a further reduction in thermal hysteresis. Reduction of the heat treatment temperatures to 600 °C promotes the R-phase formation, resulting in a transformation sequence of B2 → B2 + R-phase → B2 + R-phase + B19/B19'. In this case, the R-phase remains stable down to -100 °C, likely due to a local reduction in the $Ti_2Pd(Ni)$ precipitate spacing below a critical threshold that suppresses B19/B19' formation. In addition, increased coherent stresses and compositional variations promote a higher amount of retained austenite and significantly lower the latent heat. The martensitic transition is accompanied with a strain glass like component.

When heat treated at 500 °C the thermally induced transition likely follows a strain glass transition accompanied by weak martensitic transitions. As a result, only short-range order is achieved and the B2 austenite phase is stabilized, as evidenced by the absence of transformation peaks in the DSC, as well as minimized hysteresis as observed in the CDM compared to the other heat treatment. The transformation sequence is similar to the sample heat treated at 600 °C, but with substantial amount of macroscopically present B2 phase at -100 °C as indicated by the XRD. This can be explained by the high amount of $Ti_2Pd(Ni)$ precipitates. The spacing between the rods is in the range of 5 to 10 nm, possible smaller than the critical size necessary for the R-phase and especially the B19 phase nucleation. In addition, coherent stresses, as well as potentially strong compositional fluctuations in their vicinity lead ultimately to the formation of a strain glass. The presence and possible formation of thermal induced long range ordered R-phase and B19/B19' phase is caused by the strong heterogeneity of disorder in the system, with regions of different amounts of secondary phases and different crystallinity. The reason for the partial amorphous region is likely caused by the small difference to the determined crystallization temperature. It has to be pointed out that the strain glass state can potentially only be achieved, since a combination of different factors is present, that is formation of nanodomains with the corresponding coherent strains together with compositional fluctuations. Interestingly plate precipitates with slightly larger length scales are frequently observed in sputtered Ti-rich NiTi [55, 56]or in Ni-rich/Ti-rich TiNiCu [2]shape memory alloys and to our best knowledge and our own studies [16]there are no reports of existing remanent austenite or strain glass formation in TiNiCu based sputtered films. Although the transformation is more continuous in those alloys as well, indicated by smaller latent heats and broader DSC peaks, the local coherency strain fields induced and disorder by those precipitates are not sufficient to form a strain glass transition. The more pronounced effect of Pd doping alongside with Ni compared to Ni only surplus [24]or Cu [25] to induce strain glass behavior indicates that precipitate density, their induced stresses and compositional variations in TiNiPd all significantly contributes to stabilizing remanent austenite and promoting the strain glass formation. The influence of crystallographic compatibility on the change in thermal hysteresis needs further clarification, as the misfit of the phases is significantly influenced by the Pd content and thus varies much stronger compared to TiNiCu alloys, when compositional variation is present in the matrix by a large amount of precipitations. The change of the peak-to-peak hysteresis is mainly caused by the change in transformation sequence (Table 1), together with the reduced precipitate size, as well as the transition to strain glass behavior.

**Stress Induced Transformation**

The strong difference in the microstructure impacts the stress induced martensitic transformation and its functional fatigue as well. While crystallographic compatibility is generally considered important for shape memory behavior and its reversibility, our observations suggest that its influence may be limited in the present system. A pronounced variation in the matrix composition is observed depending on the grain location, particularly in regions of $Ti_2Pd(Ni)$ or $Ti_2Ni(Pd)$ precipitates. Pd atoms strongly influence the lattice parameter and local variations in composition lead to a continuous change in crystallographic compatibility. Lower heat treatment temperatures enhance these compositional variations, as the precipitate spacing is reduced. The growth of martensite platelets is severely hindered mainly by intergranular coherent $Ti_2Pd(Ni)$. Together with the observed compositional variation of Pd and Ti, the inhomogeneous transformation results in a sloped transformation plateau. The 800 °C sample displayed by far the smallest transformation stress, which can be explained by the large spacing of the $Ti_2Pd(Ni)$ precipitates, which does not suppress the nucleation of martensite, but only the growth is severely hindered, as can be seen from Fig. S10, SI. The large grain size, as well as the large spacing of the precipitates do not suppress the formation of dislocations, which also contributes to the large stress hysteresis compared to the other heat treatments. In contrast the significant increase in transformation stress for the 700 °C sample is likely caused by the increased volume fraction of $Ti_2Pd(Ni)$ precipitates and the probable (though not determined by TEM) smaller spacing between $Ti_2Pd(Ni)$ precipitates suppressing the nucleation of martensite. On the other hand, this leads to enhanced resistance to dislocation formation and to reduced stress hysteresis. As stress hysteresis is typically smaller with higher transformation stress the effect in hysteresis reduction is thought to be small. The transformation stress and resistance to dislocation formation is further enhanced while lowering the heat treatment temperatures causing potentially further reduced spacing of $Ti_2Pd(Ni)$ precipitates. As the R-phase is more stabilized the reversible B2 → R-phase transition is visible upon elastic loading. In addition, the superelasticity and high transformation strain is likely attributed to a partial strain glass transition. The application of stress in those cases allows for long range ordered B19/B19' martensite to form in contrast to the thermal induced transition, where an increasing invariance of B2 was observed. In the 500 °C heat treated sample, both the spacing and the size of the plate precipitates are further minimized. The formation of these nanoscale strain domains, together with the partially amorphous matrix, results in a strong suppression of martensite nucleation and thus the highest transformation stress. Stress hysteresis is potentially further reduced due to the strain glass nature. However, it should be noted that the high transformation stress leads to a greater amount of elastic strain energy being stored during the forward transformation compared to the other heat treatments, which in turn assists the reverse transformation. The high density of coherent plate precipitates effectively suppresses the dislocation formation, resulting in nearly no change in the superelastic response and further reducing hysteretic losses. In situ stress-assisted XRD show that the superelasticity and comparable transformation strain are caused by the transition of B2 nanodomains into long-range ordered martensite, as discussed above. Additional in situ XRD experiments under stress, comparing different heat treatments, are necessary to further elucidate the influence of the formed nanodomains on the transformation mechanism.

## Summary and Conclusion

In summary our study highlights the tremendous effect of the heat treatment in Ti-rich TiNiPd on the shape memory behavior, especially on the thermal and stress induced transformation and its functional fatigue. All heat treatments lead to the formation of $Ti_2Ni(Pd)$ and $Ti_2Pd(Ni)$ precipitates. The volume fraction of round $Ti_2Ni(Pd)$ intergranular and grain boundary precipitates increases with increasing heat treatment temperature and are found to be incoherent at the grain boundaries and semi coherent inside the grains. The volume fraction of $Ti_2Pd(Ni)$ precipitates has its maximum at 700 °C and is then destabilized at higher heat treatments. Most of the precipitates are found inside the grain boundaries in the form of coherent plate precipitates. It is found that the matrix composition is highly inhomogeneous depending on the position within the grains. TEM analysis reveals that the size and spacing of the $Ti_2Pd(Ni)$ precipitates is severely increased with increasing heat treatment temperature causing together with compositional variations a significant influence on the nucleation and growth of martensite. While the sample heat treated at 800 °C clearly shows a strong 1st order phase transition. A decrease in heat treatment temperature and thus the formation of smaller and denser nano-precipitates leads to an evolution of strain glass-like transformation in the material with decreasing annealing temperature. Regions with a higher density of nanoscale plate precipitates increases with decreasing heat treatment temperature as indicated by the increasing invariance of thermally induced transformation (remanent austenite) and the change of transformation path over the R-phase. This causes the 500 °C sample to show a mixture of strain glass behavior and long range ordered R-phase and B19 transformation due to the local variation of the precipitate density. In addition, the sample annealed at 500 °C is semicrystalline increasing the disorder further. In agreement with the strain glass state the presence of those nano strain domains suppress the long-range martensitic transformation as the thermal energy available during cooling is not sufficient to overcome the energy barriers introduced into the system and the phase transition is incomplete. Uniaxial tensile loading allows to overcome the local energy barriers and induce long range ordering, allowing for a near complete phase transition to B19/B19', confirming the strain glass nature of this transition. Our findings highlight the role of nano scale precipitates in continuously changing the martensitic transition to strain glass transition. The reduction of size and spacing of the $Ti_2Pd(Ni)$ precipitates leads to tremendous increase in resistance to functional fatigue, but with the costs of ductility. Further detailed studies are needed to also investigate the role of crystallographic compatibility on the transformation behavior in such systems and to distinguish between TiNiCu with plate precipitates and TiNiPd shape memory alloys. In addition, a comparison of the functional fatigue of strains glass TiNiPd caused by doping only and by precipitation only would be illuminating to understand the role of the nanodomains in terms of functional fatigue.

## Acknowledgements

The authors acknowledge funding by the German Science Foundation (DFG) through project 413288478 and the Collaborative Research Center (CRC) 1261. The authors would like to thank C. Szillus for the sample preparation using focused ion beam and W. Strohm for conducting the cantilever deflection method experiments.

## Data Sharing Policy

The data that support the findings of this study are available from the corresponding author upon request.

## Supporting Information

Additional figures are provided in the Supporting Information at the end of this document, following the References.

## References

1. K. Gall, H. Maier, Acta Materialia (2002) doi:10.1016/S1359-6454(02)00315-4
2. A. Ishida, M. Sato, Z. Gao, Acta Materialia (2014) doi:10.1016/j.actamat.2014.02.006
3. T. Waitz, V. Kazykhanov, H.P. Karnthaler, Acta Materialia (2004) doi:10.1016/j.actamat.2003.08.036
4. C. Chluba, W. Ge, T. Dankwort, C. Bechtold, R.L. de Miranda, L. Kienle, M. Wuttig, E. Quandt, Philosophical transactions. Series A, Mathematical, physical, and engineering sciences (2016) doi:10.1098/rsta.2015.0311
5. B. Malard, J. Pilch, P. Sittner, V. GARTNEROVA, R. Delville, D. Schryvers, C. Curfs, Funct. Mater. Lett. (2009) doi:10.1142/S1793604709000557
6. H. Gu, L. Bumke, C. Chluba, E. Quandt, R. D.James, Materials Today (2018) doi:10.1016/j.mattod.2017.10.002
7. R. Zarnetta, R. Takahashi, M.L. Young, A. Savan, Y. Furuya, S. Thienhaus, B. Maaß, M. Rahim, J. Frenzel, H. Brunken, Y.S. Chu, V. Srivastava, R.D. James, I. Takeuchi, G. Eggeler, A. Ludwig, Adv. Funct. Mater. (2010) doi:10.1002/adfm.200902336
8. X. Chen, V. Srivastava, V. Dabade, R.D. James, Journal of the Mechanics and Physics of Solids (2013) doi:10.1016/j.jmps.2013.08.004
9. G. Eggeler, E. Hornbogen, A. Yawny, A. Heckmann, M. Wagner, Materials Science and Engineering: A (2004) doi:10.1016/j.msea.2003.10.327

10. C. Chluba, W. Ge, R. Lima de Miranda, J. Strobel, L. Kienle, E. Quandt, M. Wuttig, Science (New York, N.Y.) (2015) doi:10.1126/science.1261164
11. J.M. Ball, R.D. James, Arch. Rational Mech. Anal. (1987) doi:10.1007/BF00281246
12. R.D. James, Z. Zhang, in *Magnetism and Structure in Functional Materials*, ed. by R. Hull, J. Parisi, R.M. Osgood, H. Warlimont, A. Planes, L. Mañosa, A. Saxena (Springer Berlin Heidelberg, Berlin, Heidelberg, 2005), p. 159
13. Y. Song, X. Chen, V. Dabade, T.W. Shield, R.D. James, Nature (2013) doi:10.1038/nature12532
14. Z. Zhang, R.D. James, S. Müller, Acta Materialia (2009) doi:10.1016/j.actamat.2009.05.034
15. A. Ishida, M. Sato, Z.Y. Gao, Intermetallics (2015) doi:10.1016/j.intermet.2014.11.011
16. L. Bumke, N. Wolff, C. Chluba, T. Dankwort, L. Kienle, E. Quandt, Shap. Mem. Superelasticity (2021) doi:10.1007/s40830-021-00354-x
17. S. Miyazaki, K. Nomura, A. Ishida, S. Kajiwara, J. Phys. IV France (1997) doi:10.1051/jp4:1997543
18. S. Inoue, K. Hori, N. Sawada, N. Nakamoto, T. Namazu, MSF (2010) doi:10.4028/www.scientific.net/MSF.638-642.2068
19. T. Sawaguchi, M. Sato, A. Ishida, Materials Science and Engineering: A (2002) doi:10.1016/S0921-5093(01)01714-2
20. M. Matsuda, D. Hashimoto, V.C. Solomon, M. Nishida, Materials Science and Engineering: A (2006) doi:10.1016/j.msea.2006.01.126
21. A.A. Klopotov, V.P. Sivokha, N.M. Matveeva, Y.A. Sazanov, Russian Physics Journal (1993) doi:10.1007/BF00559445
22. P.L. Narayana, S.-W. Kim, J.-K. Hong, N.S. Reddy, J.-T. Yeom, Metals and Materials International (2018) doi:10.1007/s12540-018-0109-4
23. S. Shimizu, Ya Xu, E. Okunishi, S. Tanaka, K. Otsuka, K. Mitose, Materials Letters (1998) doi:10.1016/S0167-577X(97)00134-1
24. S. Ren, C. Zhou, D. Xue, D. Wang, J. Zhang, X. Ding, K. Otsuka, X. Ren, Phys. Rev. B (2016) doi:10.1103/PhysRevB.94.214112
25. Yanshuang Hao, Yuanchao Ji, Zhao Zhang, Mengye Yin, Chang Liu, Hui Zhao, Kazuhiro Otsuka, Xiaobing Ren, Scripta Materialia (2019) doi:10.1016/j.scriptamat.2019.04.028
26. E. Quandt, H. Holleck, MRS Proc. (1996) doi:10.1557/PROC-459-465
27. R. Delville, D. Schryvers, Intermetallics (2010) doi:10.1016/j.intermet.2010.08.006
28. Bernhard Winzek, Eckhard Quandt, IJMR (1999) doi:10.1515/ijmr-1999-901007
29. Kengo Mitose, Tatsuhiko Ueki, Patent 5,951,793, 14 September 1999
30. P. Schlossmacher, Materials Letters (1997) doi:10.1016/S0167-577X(96)00251-0
31. R. Lima de Miranda, C. Zamponi, E. Quandt, Adv. Eng. Mater. (2013) doi:10.1002/adem.201200197
32. G.G. Stoney, C.A. Parsons, Proceedings of the Royal Society of London. Series A, Containing Papers of a Mathematical and Physical Character (1909) doi:10.1098/rspa.1909.0021
33. P. Stadelmann, MAM (2003) doi:10.1017/S1431927603012224
34. Hee Young Kim, Yimin Yuan, Tae-hyun Nam, Shuichi Miyazaki and, International Journal of Smart and Nano Materials (2011) doi:10.1080/19475411.2010.550655
35. N.M. Matveeva, Y.K. Kovneristyi, A.S. Savinov, V.P. Sivokha, V.N. Khachin, J. Phys. Colloques (1982) doi:10.1051/jphyscol:1982433
36. V.P. Sivokha, A.S. SAVVINOV, V.P. Voronin, V.N. Khachin, FIZIKA METALLOV I METALLOVEDENIE **56**, 542 (1983)
37. P. Sedmák, P. Šittner, J. Pilch, C. Curfs, Acta Materialia (2015) doi:10.1016/j.actamat.2015.04.039
38. Qingchao Tian, Jiansheng Wu, Intermetallics (2002) doi:10.1016/S0966-9795(02)00048-1
39. Shuai Ren, Chang Liu, Xuan Chen, Yanshuang Hao, Xiaobing Ren, Scripta Materialia (2020) doi:10.1016/j.scriptamat.2019.10.009
40. S. Sarkar, X. Ren, K. Otsuka, Physical review letters (2005) doi:10.1103/PhysRevLett.95.205702
41. D. Jiang, J. An, Y. Liu, Z. Ma, F. Liu, H. Yang, X. Ren, K. Yu, J. Zhang, X. Jiang, Y. Ren, L. Cui, Phys. Rev. B (2021) doi:10.1103/PhysRevB.104.024102
42. Y. Ji, X. Ding, T. Lookman, K. Otsuka, X. Ren, Phys. Rev. B (2013) doi:10.1103/PhysRevB.87.104110
43. Longjia Li, Zhongsheng Yang, Kengfeng Xu, Dingcong Cui, Lei Wang, Zhijun Wang, Junjie Li, Jincheng Wang, Feng He, Scripta Materialia (2025) doi:10.1016/j.scriptamat.2024.116449
44. Q. Liang, D. Wang, J. Zhang, Y. Ji, X. Ding, Y. Wang, X. Ren, Y. Wang, Phys. Rev. Mater. (2017) doi:10.1103/PhysRevMaterials.1.033608

45. Jian Zhang, Dezhen Xue, Xiaoying Cai, Xiangdong Ding, Xiaobing Ren, Jun Sun, Acta Materialia (2016) doi:10.1016/j.actamat.2016.08.015
46. Y. Wang, X.B. Ren, K. Otsuka, MSF (2008) doi:10.4028/www.scientific.net/MSF.583.67
47. Chao Lv, Kai Wang, Bin Wang, Jiaxing Zheng, Kaichao Zhang, Guanqi Li, Yongzhong Lai, Yu Fu, Huilong Hou, Xinqing Zhao, Acta Materialia (2024) doi:10.1016/j.actamat.2023.119598
48. Wenjia Wang, Yuanchao Ji, Minxia Fang, Dong Wang, Shuai Ren, Kazuhiro Otsuka, Yunzhi Wang, Xiaobing Ren, Acta Materialia (2022) doi:10.1016/j.actamat.2022.117618
49. S. Mathews, J. Li, Q. Su, M. Wuttig, Philosophical Magazine Letters (1999) doi:10.1080/095008399177336
50. Y. Murakami, D. Shindo, Philosophical Magazine Letters (2001) doi:10.1080/09500830110063950
51. J. Frenzel, A. Wieczorek, I. Opahle, B. Maaß, R. Drautz, G. Eggeler, Acta Materialia (2015) doi:10.1016/j.actamat.2015.02.029
52. L. Bataillard, J.E. Bidaux, R. Gotthard, Philosophical Magazine A **78**, 327 (1998)
53. J. Khalil-Allafi, A. Dlouhy, G. Eggeler, Acta Materialia (2002) doi:10.1016/S1359-6454(02)00257-4
54. J.K. Allafi, X. Ren, G. Eggeler, Acta Materialia (2002) doi:10.1016/S1359-6454(01)00385-8
55. Y. Nakata, T. Tadaki, A. Tanaka, K. Shimizu, J. Phys. IV France (1995) doi:10.1051/jp4/199558671
56. S. Kajiwara, K. Ogawa, T. Kikuchi, T. Matsunaga, S. Miyazaki, Philosophical Magazine Letters (1996) doi:10.1080/095008396179922

Supporting Information

# Effect of Ti$_2$Pd(Ni) on the Transformation Behavior in Sputtered Ti-rich TiNiPd Shape Memory Alloys

L. Bumke[1], N. Wolff[2], L. Kienle[2] and E. Quandt[1]

[1] Inorganic Functional Materials, Institute for Materials Science, Faculty of Engineering, Kiel University, Kiel, Germany

[2] Synthesis and Real Structure, Institute for Materials Science, Faculty of Engineering, Kiel University, Kiel, Germany

## EDS mapping of amorphous TiNiPd film deposited on Si wafer

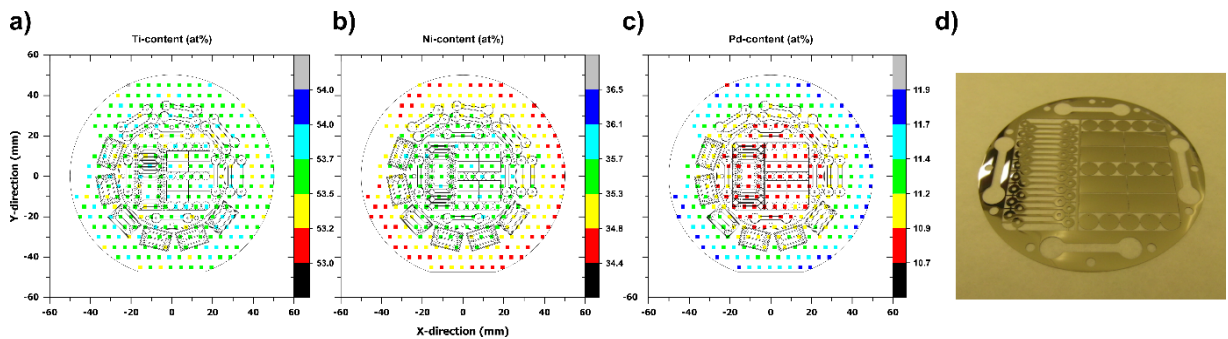

*Figure S 1 displays the compositional EDS mapping over a 4-inch Wafer of Ti (a), Ni (b), Pd(c). The CAD drawing is shown for illustration of the used area. The average Ti-content over the 4-inch wafer is: 53.6 at.% (min 51.9 at.%, max 54.2 at.%) , the Ni-content is: 35.2 at.% (min 34.4 at.%, max 36.5 at.%) and for Pd is 11.2 at.% (min 10.7 at.%, max 11.9 at.%). Within the investigated area the deviation is further minimized. In (d) exemplary a free-standing coupon of the CAD drawing is shown in the as deposited state.*

## Crystallization temperature of amorphous TiNiPd film

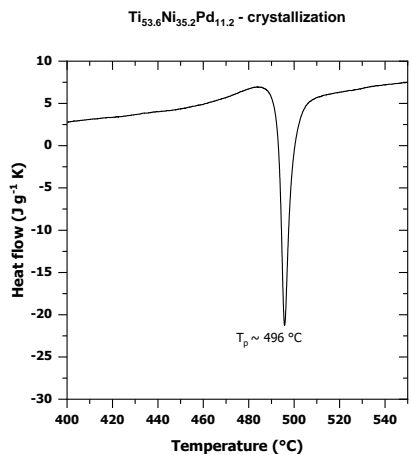

*Figure S 2 shows a heating curve of an amorphous Ti$_{53.6}$Ni$_{35.2}$Pd$_{11.2}$ film with a single exothermic peak. The crystallization temperature is determined to be 496 °C.*

## Cantilever Deflection Method for TiNiPd films

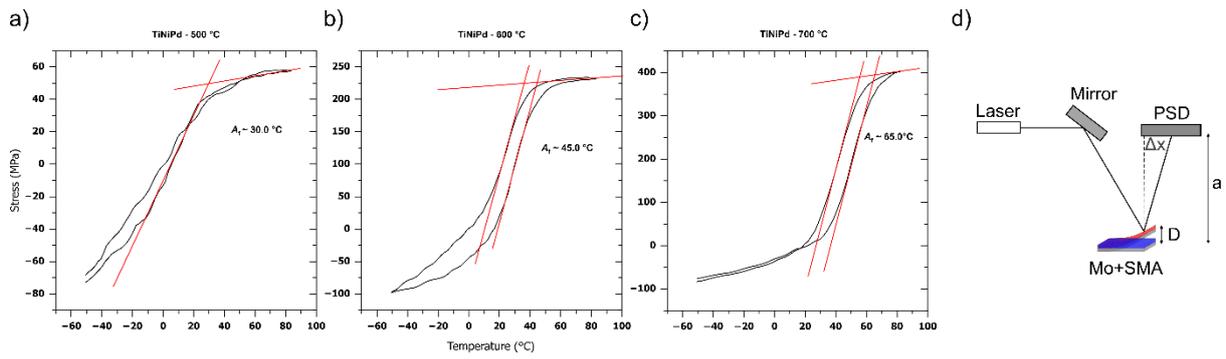

*Figure S 3 shows different stress/temperatures diagrams for the composite materials Mo/TiNiPd annealed at 500 °C (a), 600 °C (b) and 700 °C (c)for 15 minutes obtained by using the cantilever deflection method (CDM) (d). Cycles were conducted by heating and cooling between -50 °C and 80 °C using a peltier element. The laser beam is directed towards a Si-mirror mounted at the tip of the Mo/SMA composite. The actuation/stroke (D) is measured by the displacement (Δx) of the laser beam on a photo sensitive detector (PSD) which is mounted at a distance a from the cantilever.*

## Stress- Strain curves at various test temperatures for TiNiPd film heat treated at 500 °C

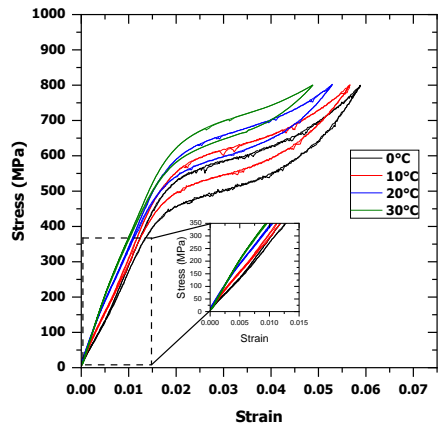

*Figure S 4 shows the temperature dependent stress strain response of a sample heat treated at 500 °C. Superelasticity is observed down to 0 °C. The enlarged region shows the temperature dependent change of slope of the elastic loading, corresponding to the B2-R-phase transition.*

## *In situ* temperature controlled XRD measurements

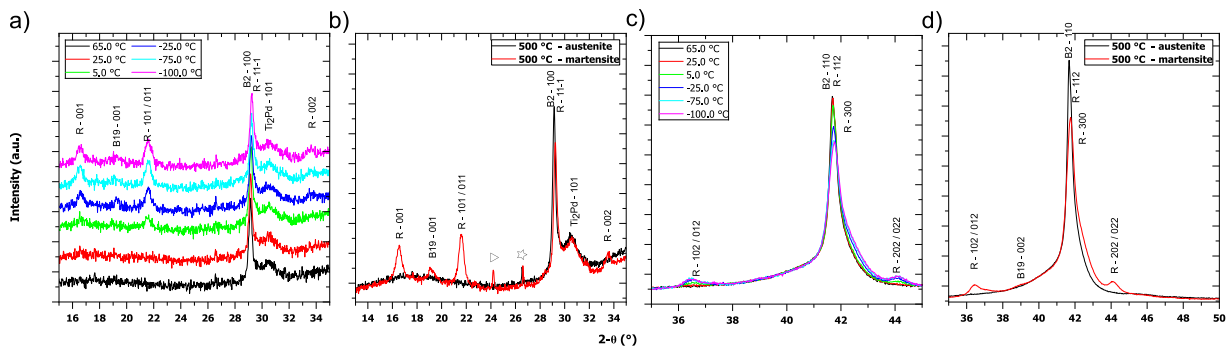

*Figure S 5 presents the temperature dependent XRD data for the 500 °C heat-treated sample. In (a) the evolution of the microstructure over the 2-θ range 15–35° at selected temperatures is depicted, while in (b) diffractograms with higher resolution at -100 °C and 90 °C are shown for the same range. The triangle and star denote reflections caused by slight condensation at -100 °C and by the thermal grease used to improve thermal contact, necessary due to slight sample bow caused by intrinsic film stress present at the 500 °C heat treated sample. The data in (c) depicts the scan range around the 110 reflection at the same test temperatures as in (a), while (d) provides a higher resolution overlay of the -100 °C and 90 °C data for enhanced clarity.*

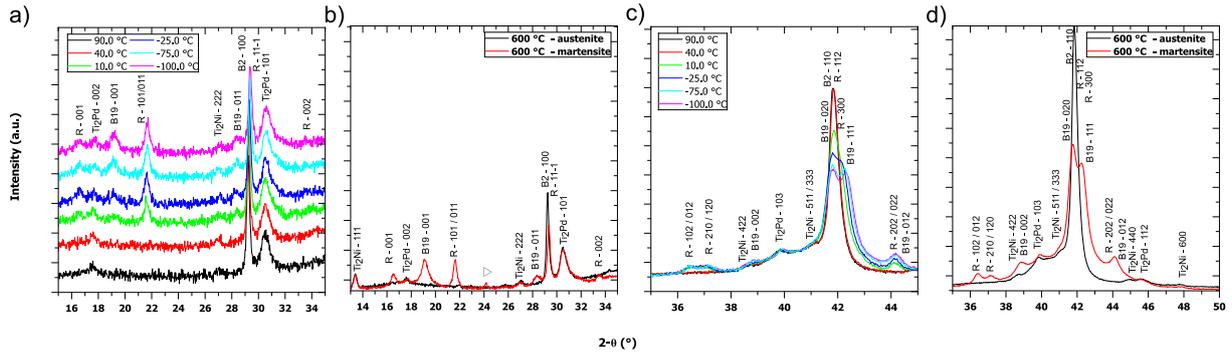

*Figure S 6 presents the temperature dependent XRD data for the 600 °C heat-treated sample. In (a) the evolution of the microstructure over the 2-θ range 15–35° at selected temperatures is depicted, while in (b) diffractograms with higher resolution at -100 °C and 90 °C are shown for the same range. The triangle denotes reflections caused by slight condensation at -100 °C. The data in (c) depicts the scan range around the 110 reflection at the same test temperatures as in (a), while (d) provides a higher resolution overlay of the -100 °C and 90 °C data for enhanced clarity.*

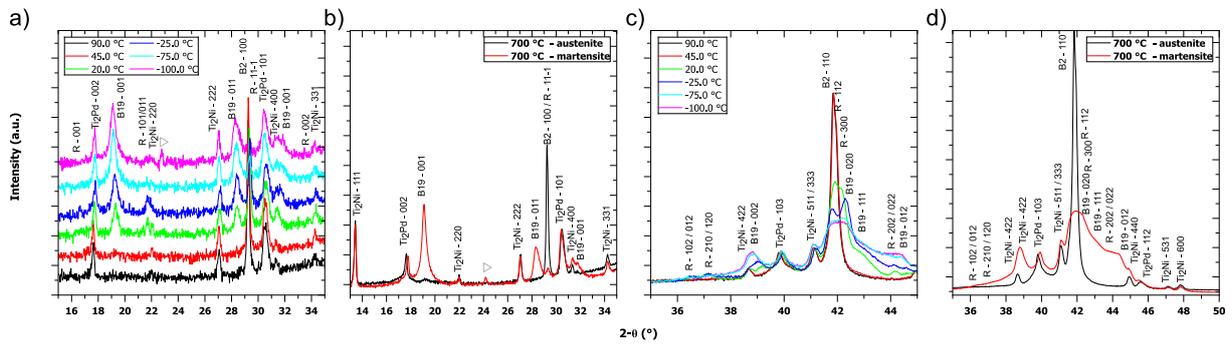

*Figure S 7 presents the temperature dependent XRD data for the 700 °C heat-treated sample. In (a) the evolution of the microstructure over the 2-θ range 15–35° at selected temperatures is depicted, while in (b) diffractograms with higher resolution at -100 °C and 90 °C are shown for the same range. The triangle denotes reflections caused by slight condensation at -100 °C. The data in (c) depicts the scan range around the 110 reflection at the same test temperatures as in (a), while (d) provides a higher resolution overlay of the -100 °C and 90 °C data for enhanced clarity.*

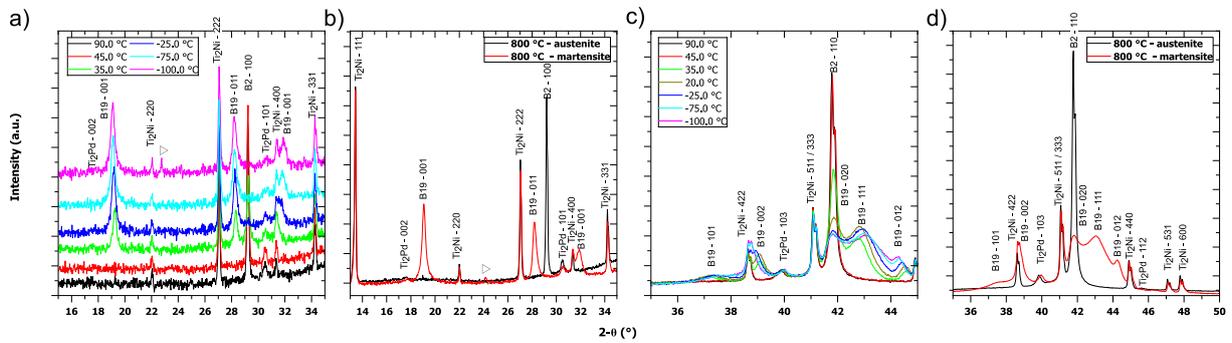

*Figure S 8 presents the temperature dependent XRD data for the 800 °C heat-treated sample. In (a) the evolution of the microstructure over the 2-θ range 15–35° at selected temperatures is depicted, while in (b) diffractograms with higher resolution at -100 °C and 90 °C are shown for the same range. The triangle denotes reflections caused by slight condensation at -100 °C. The data in (c) depicts the scan range around the 110 reflection at the same test temperatures as in (a), while (d) provides a higher resolution overlay of the -100 °C and 90 °C data for enhanced clarity.*

## Grain sizes revealed by wet chemical etching

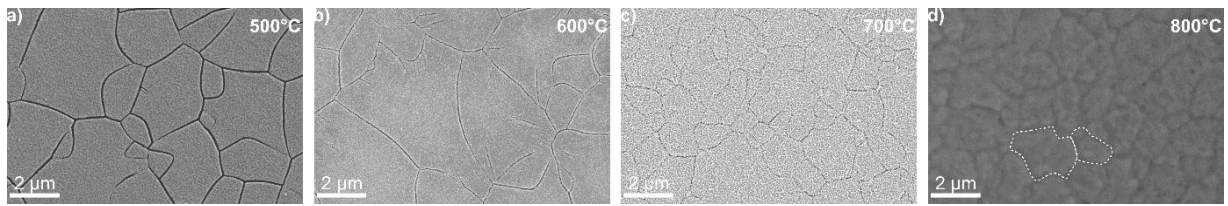

*Figure S 9 shows the different grain sizes after wet chemical etching for heat treatment temperature of 500 °C (a), 600 °C (b), 700 °C (c) and 800 °C (d). The grain size for the samples annealed at 500 °C and 600 °C is slightly larger compared to the samples annealed at 700 °C and 800 °C with 1-2 µm large grains. Two grains are framed for better visibility in (d).*

## Beam heating induced phase transition for TiNiPd sample heat treated at 800 °C

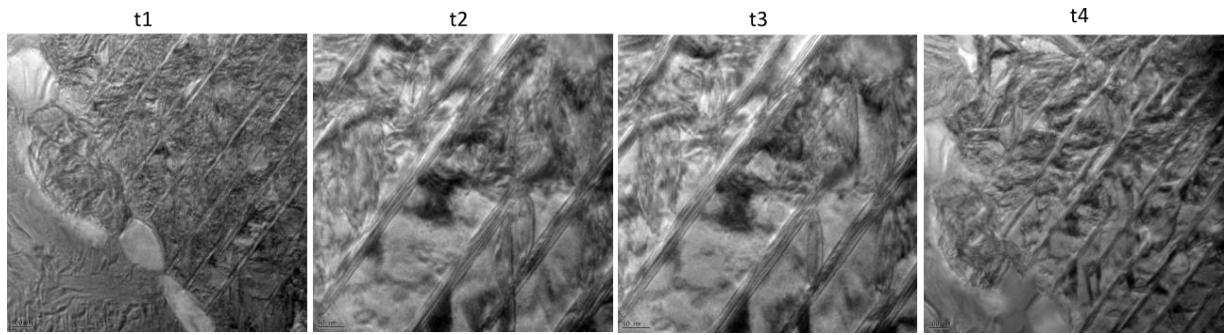

*Figure S 10 illustrates the microstructure evolution under prolonged beam exposure. Left (t1) displays an overview image of the microstructure directly at the beginning of beam exposure. At higher magnification (t2) martensite laths confined between the $Ti_2Pd(Ni)$ rods are visible. With continued beam exposure and heating (t3), these martensite variants gradually vanish between the $Ti_2Pd(Ni)$ platelets. The $Ti_2Pd(Ni)$ rods seem to impede the growth of the martensite lath and non-uniform transformation. At t4 the overview image clearly demonstrates the phase change from martensite to mostly austenite.*

## HRTEM of $Ti_2Pd(Ni)$ and $Ti_2Ni(Pd)$ preciptates for TiNiPd sample heat treated at 800 °C.

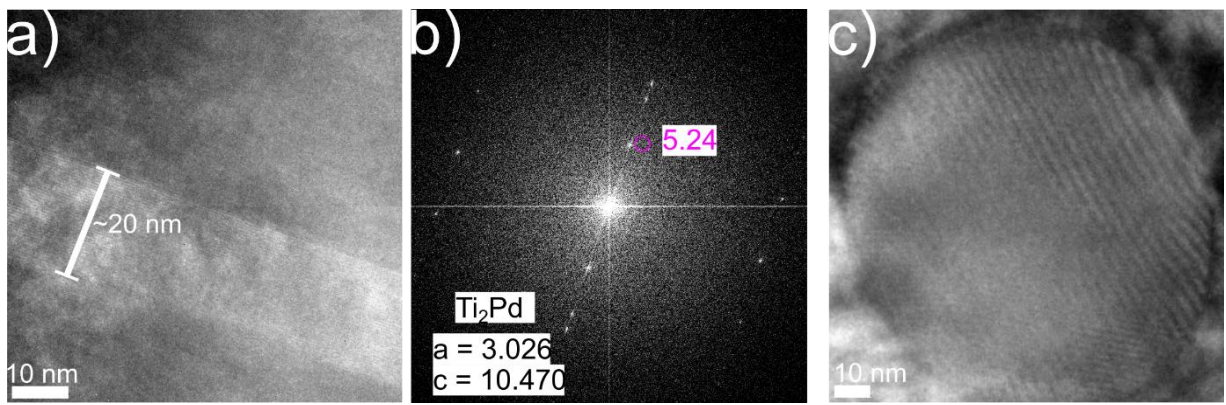

*Figure S 11 displays an intergranular $Ti_2Pd(Ni)$ rod precipitates with lattice parameters of a=3.026 Å and c=10.470 Å as determined by the FFT (b). In (c) an intergranular $Ti_2Ni(Pd)$ precipitate is positioned on top of a TiNiPd-B2 grain exhibiting lattice strain as evident from the formation of Moiré fringes on intergranular $Ti_2Ni(Pd)$ caused by the lattice mismatch with TiNiPd B2 phase.*

## TEM EDS mapping for TiNiPd sample heat treated at 800 °C

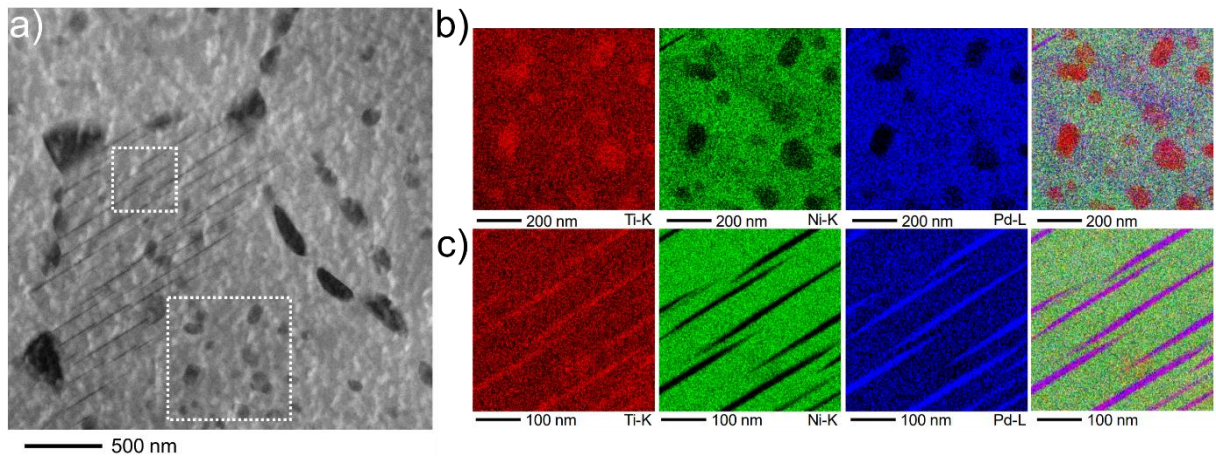

*Figure S 12 displays additional EDS maps for the microstructure observed in (a). In (b) single elemental maps of Ti-K (red), Ni-K (green), Pd-L (blue) and superposition image recorded by EDS of intergranular $Ti_2Ni(Pd)$ are depicted. In (c) single elemental maps of Ti-K (red), Ni-K (green), Pd-L (blue) and superposition image recorded by EDS of intergranular $Ti_2Pd(Ni)$ are depicted.*

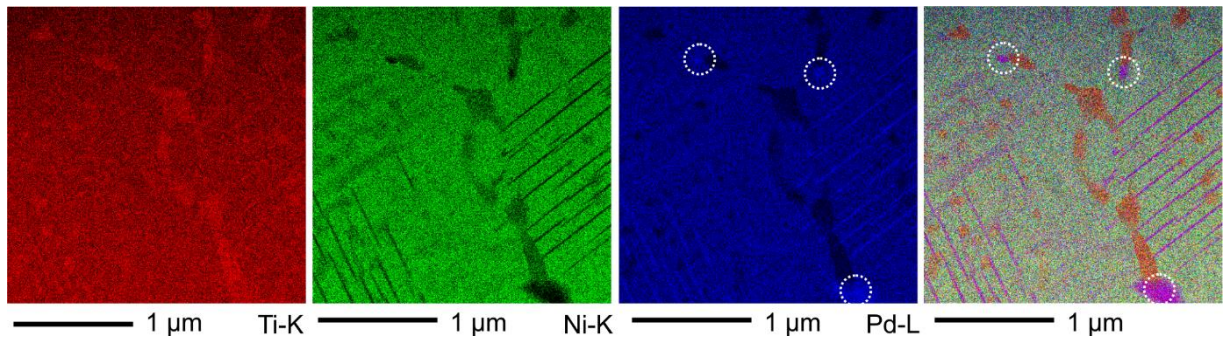

*Figure S 13 displays additional EDS maps showing the elemental maps of Ti-K (red), Ni-K (green), Pd-L (blue) and superposition image. The presence of $Ti_2Pd(Ni)$ precipitates at the grain boundaries is highlighted by the dashed circles.*

*Table S 1 Chemical composition of intergranular $Ti_2Ni(Pd)$, $Ti_2Pd(Ni)$ precipitates and grain boundary $Ti_2Ni(Pd)$, $Ti_2Pd(Ni)$ precipitates. In addition, the composition in between $Ti_2Ni(Pd)$ and $Ti_2Pd(Ni)$ was determined from the microstructure given in Fig S12. Standard deviations are provided when data from more than two precipitates are analyzed.*

| Sample ID | Ti (at.%) | Ni (at.%) | Pd (at.%) |
| --- | --- | --- | --- |
| $Ti_2Ni(Pd)$ – IG (5 precipitates) | 55.7±1.9 | 33.9±0.8 | 10.5±1.2 |
| $Ti_2Ni(Pd)$ – GB (4 precipitates) | 62.6±0.2 | 30.7±0.1 | 6.7±0.1 |
| $Ti_2Pd(Ni)$ – IG (6 precipitates) | 57.5±0.8 | 17.4±0.9 | 25.1±0.4 |
| $Ti_2Pd(Ni)$ – GB (2 precipitates) | 60.5 | 14.1 | 25.4 |
| $Ti_2Ni(Pd)$ – Area-1 | 48.6 | 37.7 | 13.7 |
| $Ti_2Ni(Pd)$ – Area-2 | 47.1 | 39.9 | 12.9 |
| $Ti_2Pd(Ni)$ – Area | 45.8 | 41.5 | 12.7 |